# Precise Balancing of Viscous and Radiation Forces on a Particle in Liquid-Filled Photonic Bandgap Fiber

T. G. Euser[*], M. K. Garbos, J. S. Y. Chen, and P. St.J. Russell

Max Planck Institute for the Science of Light, Günther-Scharowsky-Str. 1/Bau 24, 91058 Erlangen, Germany    www.pcfibre.com

* To whom correspondence should be addressed. E-mail: tijmen.euser@mpl.mpg.de

**Abstract**

**It is shown that, in the liquid-filled hollow core of a single-mode photonic crystal fiber, a micron-sized particle can be held stably against a fluidic counter-flow using radiation pressure, and moved to and fro (over 10s of cm) by ramping the laser power up and down. The results represent a major advance over previous work on particle transport in optically multimode liquid-filled fibers, in which the fluctuating transverse field pattern renders the radiation and trapping forces unpredictable. The counter-flowing liquid can be loaded with sequences of chemicals in precisely controlled concentrations and doses, making possible studies of single particles, vesicles or cells.**

The synthesis, measurement and manipulation of micron-sized objects is of great importance in many fields, from catalysis to cell biology and cancer research, from quantum dots to solar cells, from paint design to colloidal chemistry. Optical tweezering, combined with microfluidics (*1-3*), is a major field that has been used, for example, to size-sort dielectric particles (*4*) and manipulate mammalian cells (*5*). Radiation pressure has also been used to guide particles along hollow photonic crystal fibres (*6,7*) and multi-mode liquid-filled fibre cores (*8*), and to propel atoms along hollow capillaries (*9*).

Here we report for the first time that a micron-sized particle, trapped in the liquid-filled core of a single-mode photonic crystal fibre, can be held motionless against a fluidic counter-flow or moved to and fro by adjusting the laser power. Unlike in previous experiments, the particle sits in the centre of a single guided optical mode, feeling the full force of the laser light. The optical and viscous forces can thus be precisely balanced, and the low optical attenuation, in combination with the absence of diffraction, allows particles to be moved at will over extended distances (tens of cm), which compares favourably with the 1 mm distances so far achieved with non-diffracting Bessel beams (*10*). The mechanical flexibility of optical fibre, together with low optical bend-loss, makes it feasible to guide particles along reconfigurable curved trajectories, something that is impossible using free-space beams. There are many potential applications for this new system. Very small flows (tens of pL s$^{-1}$) can be counter-balanced at moderate optical powers (tens of mW), allowing for example cells or vesicles to be held motionless while drugs or chemicals flow past in highly controlled quantities. Side-illumination through the transparent cladding would allow photo-activation of, e.g., novel anti-cancer compounds in the liquid (*11*), and fluorescence could be monitored either through the cladding or along the guiding core.

The unique combination of single mode guidance and low loss also makes possible precise measurements of the drag forces acting on single particles – as studied in this article. Photonic band gap confinement permits low loss optical guidance even when the refractive index in the liquid is lower than in the cladding. If only the hollow core is filled with liquid, leaving the cladding holes empty, the large index contrast between cladding and core allows guidance by total internal reflection (*8*). However, the result is a multimode waveguide in which the transverse intensity pattern is a difficult-to-control, axially varying, superposition of many modes. The evanescent edging field of a single guided optical mode may also be used to propel particles over short distances (~0.1 mm) on planar waveguides (*12, 13*) and in Si slot-waveguides (*14*). This approach has the disadvantage that the transverse optical field decays exponentially from the surface, making stable optical trapping difficult. Furthermore, the particles are guided very close to the waveguide surface, resulting in asymmetric drag forces, in contrast to the experiments reported here, where the particle is held in the middle of the flowing liquid.

**Experimental arrangements**

The photonic crystal fibre used had a core diameter of 17 μm (see Fig. 1) and was designed, following known scaling laws (*15, 16*), for single-mode guidance at wavelength 1064 nm when filled with deuterium oxide. $D_2O$ was used because its absorption (0.04 dB cm$^{-1}$) at the trapping wavelength 1064 nm is 15 times lower than that of $H_2O$ (0.6 dB cm$^{-1}$), minimising the effects of laser heating. Custom-designed liquid cells were used in the filling process. Coupling to the guided mode was optimized using an objective lens (4× 0.1 NA) that matched the numerical aperture of the guided mode in the liquid-filled fibre. Launch efficiencies of ~89% into the fundamental core mode were achieved. Fig. 1C shows the measured near-field mode profile at the output face of an 11 cm long piece of $D_2O$-filled fibre. Robust single-mode mode guidance was obtained over the wavelength range 790 nm to 1140 nm. Using a cut-back technique, the loss was measured to be 0.05 dB cm$^{-1}$ at 1064 nm – only slightly higher than the absorption of $D_2O$.

An 11 cm length of liquid-filled fibre was placed horizontally on a glass plate, its input face oriented parallel to a vertical glass window (100 μm thick) and immersed in a $D_2O$ droplet (Fig. 1D). The end-face of the fibre was enclosed in a pressure cell and imaged through an optical window using camera CCD4 (Fig. 1E). The pressure applied to the system could be adjusted over the range ±2 kPa by raising or lowering the $D_2O$ reservoir. The set-up allowed the peak flow velocity to be accurately adjusted over the range ±266 μm s$^{-1}$.

The light from a continuous wave Nd:YAG laser (1064 nm) was divided at a beam-splitter into guidance and loading beams (Fig. 1D). The loading beam was focused by a long working-distance (100× 1.1 NA) water immersion objective, forming a conventional single-beam optical tweezer trap (*17*). Cameras CCD1 and CCD2, monitoring the input face from orthogonal directions, allowed three-dimensional control of particle position.

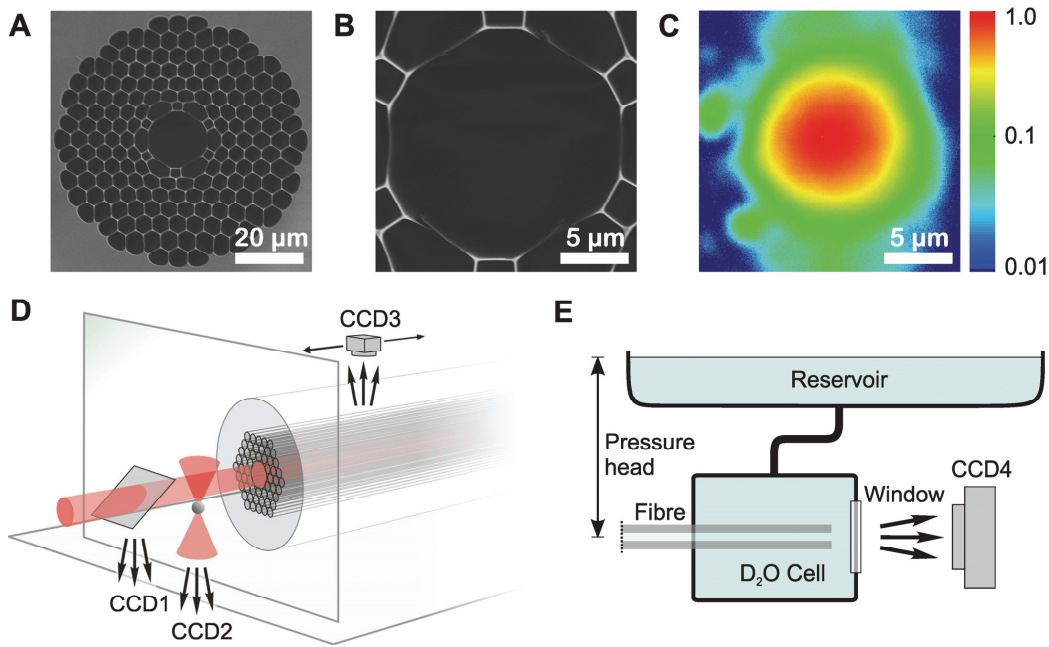

**Fig. 1.** (**A**) SEM of the cross-section of the HC-PCF with inter-hole spacing $\Lambda = 4.7\pm0.1$ µm and core diameter 17.1±0.3 µm; (**B**) magnified image of the core structure; (**C**) mode intensity profile measured at the output of a 11 cm piece of $D_2O$-filled fibre at a wavelength of 1064 nm (same scale as (**B**)); (**D**) arrangement for loading and launching particles into the fibre and monitoring them while inside; the fibre end-face is immersed in a drop of liquid at the vertex of two lass slides; (**E**) pressure cell and reservoir at output end of fibre.

A small amount of dilute silica sol was added to the $D_2O$ droplet at the fibre input face. A single particle was selected from those in the droplet, trapped by the loading beam, and moved to the entrance of the fibre core. Photographs of this process, as seen by camera CCD1, are shown in Fig. 2A-C. Once the particle had reached the core entrance, the loading beam was blocked, and the horizontal guidance beam was used to push it into the core. The image from CCD2 in Fig. 2D shows the particle trapped just outside the core entrance by a combination of fluid counter-flow and radiation force. Upon increasing the optical power, the particle is pushed into the core, after which the transmitted power drops by ~40%. Once securely trapped inside the fibre, the particle could be moved to and fro by adjusting the laser power and the fluid counter-flow.

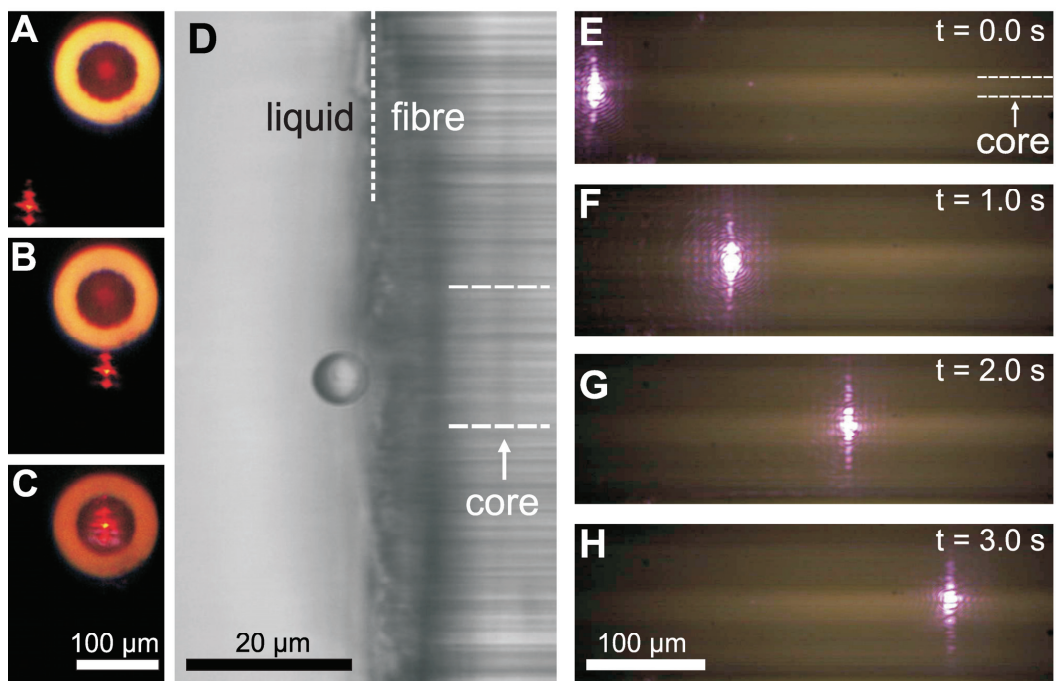

**Fig. 2**. Loading, launching and guidance of a particle (diameter 6 μm). (**A**) to (**C**) tweezering a particle up to the entrance to the core; (**D**) side-view of the particle held at the entrance to the core by optical forces balanced against counter-flow of liquid from the core. Fluid flow and gravity push the particle slightly below core centre. While in this position the particle could be seen to revolve under the action of imbalanced viscous forces; (**E**) to (**H**) side-scattering patterns imaged through the cladding of the fibre, photographed at 1 s intervals.

**Theory**

The Reynolds number ($= 2\rho \bar{V} R / \eta$, where $\rho$ is the liquid density, $\bar{V}$ the average fluid velocity, $R = 8.5$ μm is the core radius and $\eta$ the dynamic viscosity) has the value 0.0015 for $\bar{V} = 100$ μm s$^{-1}$ (typical in the experiments), $\rho = 1106$ kg m$^{-3}$ and $\eta = 0.00125$ N s m$^{-2}$, indicating laminar flow. This allows us to use Hagen-Poiseuille theory for an incompressible fluid. Theory shows that the flow rate is not noticeably affected by opposing particle motion under our experimental conditions ($\zeta = a/R < 0.4$ where $a$ is the particle radius).

The viscous drag force on a particle being pushed through a constrained counter-flow is complicated to calculate, requiring numerical methods (*18*). Two limiting regimes can be identified. The first arises when the flow is zero and the particle proceeds at constant speed under the action of the optical force, and the second when the particle is held stationary against the flow by the optical force. In the general case, the net drag force is the sum of these two, and can be written:

$$F_{net} = -6\pi \eta a \left( V_p K_1(\zeta) - V_{max} K_2(\zeta) \right) \quad (1)$$

where $V_{max} = (R^2 / 4\eta) \cdot dp/dz$ is the fluid velocity in the centre of the core, $V_p$ the particle velocity, $\eta$ is the viscosity and $p$ is the pressure at point $z$ along the fibre. Using data from Quddus (*18*), the numerically-evaluated correction factors $K_1$ and $K_2$ can be represented (<1% error) by the following polynomials (19):

$$K_1(\zeta) = 1.79 - 13.6\zeta + 108\zeta^2 - 273\zeta^3 + 290\zeta^4$$
$$K_2(\zeta) = 1.48 - 7.51\zeta + 67.1\zeta^2 - 166\zeta^3 + 180\zeta^4. \tag{2}$$

The optical forces are more difficult to estimate in a waveguide geometry, where it is not clear that the assumptions of the standard ray-optics approach (*20*) are valid. Nevertheless, in order to provide a basis for comparison, we describe in the Supporting Online Material (*19*) an analysis where the light guided in the core is represented by a bundle of rays travelling parallel to the axis, with intensities following the $J_0^2(j_{01}\, r/R)$ shape expected for the fundamental mode ($j_{01}$ is the first zero of the $J_0$ Bessel function). The momentum transferred to the particle is calculated for each ray, and then integrated over all rays. The propulsive force for a particle with radius $a$ sitting in the centre of the beam can be represented by the polynomial:

$$F_p = 0.214 - 1.65\,a + 13.4\,a^2 - 1.64\,a^3 \quad \text{pN W}^{-1} \tag{3}$$

with $a$ in µm and $R = 8.5$ µm. The refractive indices of silica and $D_2O$ were taken to be 1.45 and 1.33 respectively. Under the same conditions, the lateral restoring force per unit displacement from centre is:

$$k_T = 0.0387\,a - 0.106\,a^2 + 1.58\,a^3 - 0.0653\,a^4 - 0.0181\,a^5 \quad \text{pN W}^{-1}\mu\text{m}^{-1} \tag{4}$$

Using these expressions, it is interesting to calculate the downward displacement due to gravity of a silica particle optically held in a horizontal $D_2O$-filled fibre. At 50 mW laser power, it is 0.5 µm for $a = 1$ µm and 0.62 µm for $a = 3$ µm, taking the density of silica as 2000 kg m$^{-3}$. If flowing, the liquid will provide an extra trapping force, further reducing this already small deflection.

**Results**

Measurements were made for two different drag regimes. In the first, the particle velocity in the absence of any flow was measured via side-scattering using camera CCD3. A sequence of typical photographs, taken at 1 second intervals, is shown in Fig. 2E-H. The velocities are plotted against optical power in Fig. 3A, showing that the power-dependent optical transport velocity $dV_p/dP_{opt}$ lies in the range of 0.5 to 1 mm s$^{-1}$ W$^{-1}$ for particle radii between 1 and 3 μm. In the second experiment, the reservoir was positioned so as to create a continuous flow against the direction of the light, and the optical power was adjusted so that the particle remained stationary in the laboratory frame. This was repeated for a range of different pressure gradients, and for two different sphere sizes. The results show a linear relationship (with slopes $d^2P_H/dz.dP_{opt}$ from 0.7 to 1 kPa cm$^{-1}$ W$^{-1}$, depending on particle size) between pressure gradient and the optical power needed to keep the particle stationary (Fig. 3B).

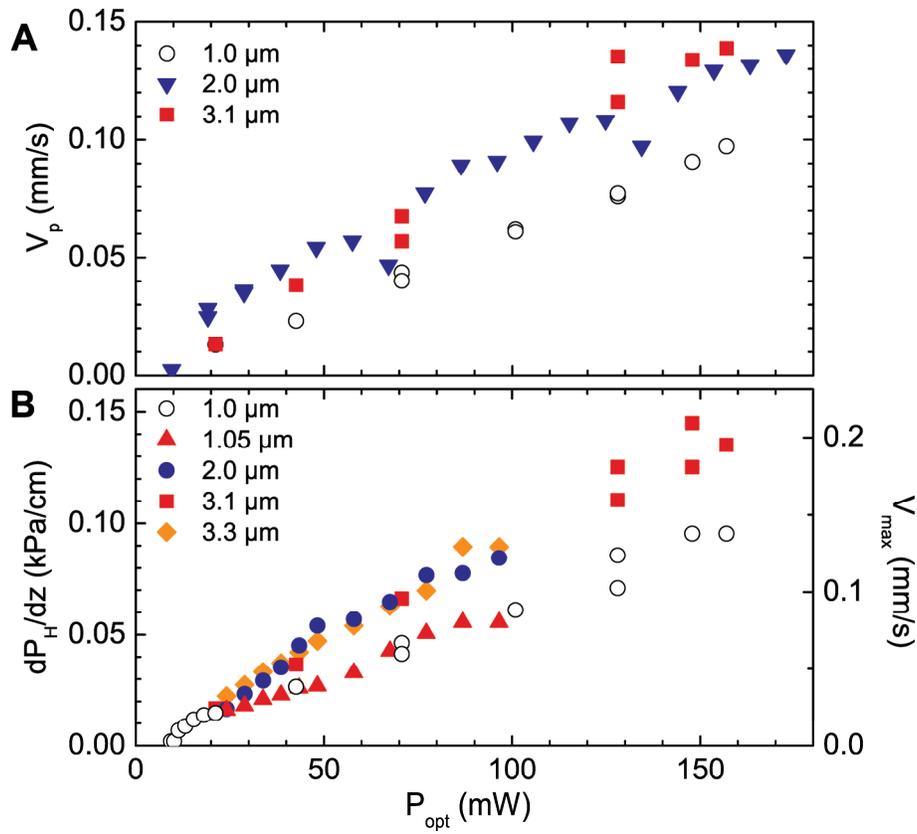

**Fig. 3.** Experimental data for the limiting cases of stationary fluid and stationary particle. (**A**) Particle velocity $V_p$ versus launched optical power $P_{opt}$ for three different particle sizes (zero liquid flow). The relationship is approximately linear. At low powers the transverse trapping strength is weak, causing the particle to move closer to the wall and lowering $V_p$. (**B**) optical power needed to hold a silica sphere in a stable position against the fluid flow induced by pressure gradient $dP_H/dz$ for five different particle radii. The right-hand axis shows the velocity $V_{max}$ in the centre of the flow. Once again the relationship is linear, after an initial switch-on power when the particle is lifted into the liquid by the light.

**Analysis**

Comparisons with the predictions of theory are shown in Table 1. At zero flow, theory consistently over-estimates, by a factor of ~3 for the larger particles and ~1.8 for the smaller ones, the power needed to reach a given particle velocity. The disagreement is slightly larger for the pressure gradient required to make the particles stationary; in this case theory over-estimates the power required by factors of ~2.5 and ~4 for small and large particles. We suggest that this disagreement may be due to the waveguide geometry, which restricts the free propagation of rays escaping from the particle. In particular, some of the rays will lie within the capture angle of the waveguide mode. The coherent sum over all such rays is unlikely to have a phase or amplitude profile that matches the guided mode, which will result in less transmitted light in the guided mode and a stronger propulsive force. Another possible contributing factor is the presence of Mie resonances, which could also increase the propulsive optical force for a given power. Both these effects will also be more pronounced for larger particles. A full explanation of this must, however, await the results of an on-going analysis of the complex scattering behaviour of a particle in hollow-core photonic crystal fibre.

**Outlook**

The system described offers fresh possibilities for studying the forces acting on particles in microfluidic channels. For example, if a trapped particle is pushed sideways using a laterally-focused laser beam (which can be delivered through the cladding (*21*) ), the imbalance of viscous drag on opposite sides will cause it to spin, enhancing chemical reactions at the particle surface. Such effects have already been observed (see above) when the particle is being launched into the fibre.

By loading the flowing liquid with chemicals in sequence (and perhaps activating them photolytically by side-illumination), an optically-trapped particle could be coated with multiple layers of different materials in a highly controlled manner, the reaction being monitored using in- or through-fibre spectroscopy. This technique could have uses, for example, in the development and optimisation of colloid-based catalysts. The liquid-filled fibre can be viewed as a miniature "riser-downer" reactor, the chemicals needed for synthesis flowing one way, while the particles flow the other. It should also be possible to implement this technique in the gas phase by filling the fibre with suitable gases or vapours.

In biomedical research, while a cell is optically held against a counter flow, minute amounts of drugs (perhaps photo-activated) could be loaded into the liquid. The highly controlled micro-environment could then be used to study the effectiveness of chemical therapy at the single cell level. Since the refractive index of cancer cells is higher than that of healthy ones (1.37), it may even be possible to distinguish them by their larger velocities under optical propulsion through the fluid.

The spectrally broad window of transmission would allow in situ spectroscopic analysis of particles, cells or quantum dots trapped in the waveguide. Mie resonances, if present, could be used as an indicator of particle size, and micro-Raman techniques could be used to detect changes in the cell membrane or chemical structure.

Finally, the system could be used as a flexible opto-fluidic interconnect for transporting particles or cells from one microfluidic circuit to another.

**Table 1.** Comparison of theory and experiment.

| $a$ (μm) | Exp/Th | 1.00 | 1.05 | 2.00 | 2.00 | 3.10 | 3.30 |
|---|---|---|---|---|---|---|---|
| $d^2P_H/dz.dP_{opt}$ [kPa cm$^{-1}$ W$^{-1}$] | Exp | 0.62 | 0.61 | - | 0.94 | 0.91 | 0.95 |
| | Th | 0.24 | 0.24 | - | 0.30 | 0.25 | 0.24 |
| Correction factor | | 2.6 | 2.5 | - | 3.1 | 3.6 | 4.0 |
| | | | | | | | |
| $dV_p/dP_{opt}$ [mm J$^{-1}$] | Exp | 0.62 | - | 0.70 | - | 1.00 | - |
| | Th | 0.34 | - | 0.42 | - | 0.33 | - |
| Correction factor | | 1.8 | - | 1.7 | - | 3.0 | - |
| | | | | | | | |
| $dV_{max}/dV_p$ | Exp | 1.44 | - | - | - | 1.31 | - |
| $dV_{max}/dV_p = K_1/K_2$ | Th | 1.00 | - | - | - | 0.91 | - |
| Correction factor | | 1.44 | - | - | - | 1.44 | - |

The theoretical values are calculated using Eqs. (1), (2) and (3) (see Supporting Online Material for more details (*21*)). The correction factors are the values needed to make theory and experiment agree. The last two rows compare the experimental results for the two limiting cases of viscous drag with the theoretical predictions of Quddus (*19*); for identical optical power the propulsive force in each case is the same.

# Appendix

### S1.  Viscous drag forces

Two limiting regimes of viscous drag force can be identified for a particle being pushed through a constrained counter-flow in a narrow pipe. The first arises when the particle is held stationary against the flow by the optical force, and the second when the flow is zero and the particle proceeds at constant speed under the action of the optical force. In the general case, the net drag force is the sum of these two, and can be written as (*19*):

$$F_{net} = -6\pi \eta a \left( V_p K_1(\zeta) - V_{max} K_2(\zeta) \right), \quad \zeta = a/R \tag{s1}$$

where $V_{max} = (R^2/4\eta) \cdot dp/dz$ is the fluid velocity in the centre of the core (calculated assuming Poiseuille flow), $V_p$ the particle velocity, $\eta$ is the viscosity and $p$ is the pressure at point $z$ along the fibre. Using data from Quddus, the numerically-evaluated correction factors $K_1$ and $K_2$ can be represented (<1% error) by the following polynomials:

$$\begin{aligned} K_1(\zeta) &= 1.79 - 13.6\zeta + 108\zeta^2 - 273\zeta^3 + 290\zeta^4 \\ K_2(\zeta) &= 1.48 - 7.51\zeta + 67.1\zeta^2 - 166\zeta^3 + 180\zeta^4. \end{aligned} \tag{s2}$$

## S2. Optical forces based on ray picture

We represent the light guided in the core as a bundle of rays travelling parallel to the axis, with intensities following the $J_0^2(j_{01}\, r/R)$ shape expected for the fundamental mode ($j_{01}$ is the first zero of the $J_0$ Bessel function). The momentum transferred to the particle is calculated for each ray, and then integrated over all rays. A transmitted ray is generated every time the ray inside the sphere strikes the boundary. Summing over all these rays, and including the incident ray, yields the element of force:

$$dF_z = \frac{n_L dP}{c}\left(1 + R\cos 2\alpha - \frac{T^2\left(\cos(2\alpha - 2\theta) + R\cos 2\alpha\right)}{1 + R^2 + 2R\cos 2\theta}\right)$$

$$dF_{xy} = \frac{n_L dP}{c}\left(R\sin 2\alpha - \frac{T^2\left(\sin(2\alpha - 2\theta) + R\sin 2\alpha\right)}{1 + R^2 + 2R\cos 2\theta}\right)$$

(s3)

where $dP$ is the power in the incident ray and $n_L$ the refractive index of the liquid. The Fresnel power coefficients $R$ and $T$ take the usual forms, and the angle $\theta$ is given by Snell's law for angle of incidence $\alpha$ (see Fig. S1A). The transverse force $dF_{xy}$ points in the plane of the rays in Fig. S1A.

To find the total forces acting on the sphere, these elemental forces are integrated over all incident rays. For small index contrast the Fresnel coefficients for $s$ and $p$ polarisation are very similar, so we have followed Ashkin in taking the mean of the two forces. After numerical integration, the total propulsive force for a particle with radius $a$ sitting in the centre of the core (radius $R = 8.5$ μm) works out to be:

$$F_p(a) = 0.214 - 1.65\, a + 13.4\, a^2 - 1.64\, a^3 \quad \text{pN W}^{-1} \qquad \text{(s4)}$$

with *a* in μm. The refractive indices of silica and D$_2$O were taken to be 1.45 and 1.33 respectively. Under the same conditions, the lateral restoring force per unit displacement from centre is:

$$k_\text{T} = 0.0387\,a - 0.106\,a^2 + 1.58\,a^3 - 0.0653\,a^4 - 0.0181\,a^5 \quad \text{pN W}^{-1}\,\mu\text{m}^{-1} \qquad \text{(s5)}$$

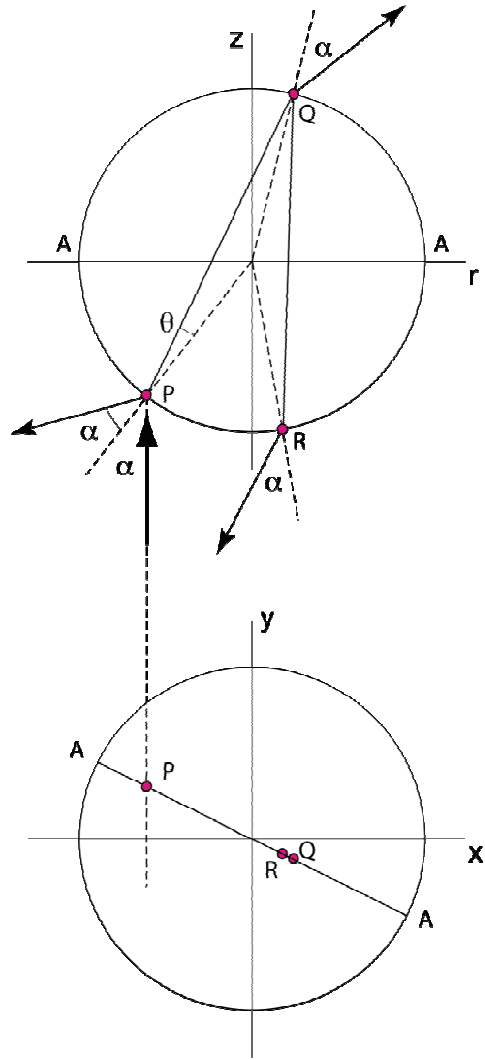

**Fig. S1.** Ray path inside spherical particle. The upper sketch shows a plan-view of the rays propagating in the plane AA, up to the second internal reflection. The lower figure is the particle as seen by the incident guided mode – the plane AA is tilted in the (x,y) transverse plane. The total force on the particle is obtained by integrating over all incident points P.

## S3. Balancing viscous and optical forces

The balance between viscous forces and the *z*-component of the optical force is obtained by equating (s4) and (s1) for the two limiting cases of stationary particle (non-zero flow):

$$F_p(a) P_{opt} = 6\pi \eta a |V_{max}| K_2(a/R) \qquad (s6)$$

and zero flow (moving particle):

$$F_p(a) P_{opt} = 6\pi \eta a |V_p| K_1(a/R) \qquad (s7)$$

with laser power $P_{opt}$ in W. $V_{max}$ is the flow velocity in the middle of the core, which can be calculated from the pressure gradient assuming Poiseuille flow (see main text of article).